\documentclass{PoS}

\newcommand{\beqa}{\begin{eqnarray}}
\newcommand{\eeqa}{\end{eqnarray}}

\title{Extension of the HAL QCD approach to inelastic and multi-particle scatterings in lattice QCD}

\ShortTitle{Extension of the HAL QCD approach to inelastic and multi-particle scatterings in lattice QCD}

\author{\speaker{Sinya AOKI}  \\
        Yukawa Institute for Theoretical Physics, Kyoto University, Kitashirakawa Oiwakecho, Sakyo-ku, Kyoto 606-8502, Japan\\
        E-mail: \email{saoki@yukawa.kyoto-u.ac.jp}}
\author{for HAL QCD Collaboration\\
\begin{center}
\includegraphics[width=0.35\textwidth]{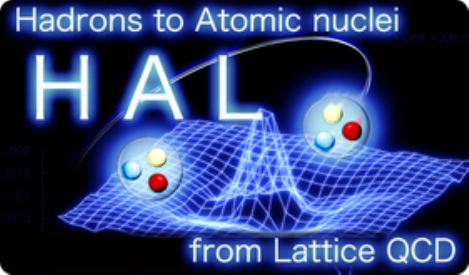}
\end{center}
}

\abstract{We extend the HAL QCD approach, with which potentials between two hadrons can be obtained in QCD at energy  below inelastic thresholds,  to inelastic and multi-particle scatterings.
We first derive  asymptotic behaviors of the Nambu-Bethe-Salpeter (NBS) wave function at large space separations for systems with more than 2 particles, in terms of the one-shell $T$-matrix consrainted by the unitarity of quantum field theories.
We show that its asymptotic behavior contains phase shifts and mixing angles of $n$ particle scatterings.
This property is one of the essential ingredients of the HAL QCD scheme to define "potential" from the NBS wave function in quantum field theories such as QCD. 
We next construct energy independent but non-local potentials above  inelastic thresholds,
in terms of these NBS wave functions. We demonstrate an existence of energy-independent coupled channel potentials with a non-relativistic approximation, where momenta of all particles are small compared with their own masses. 
Combining these two results, we can employ  the HAL QCD approach also to investigate inelastic and multi-particle scatterings.
 }

\FullConference{31st International Symposium on Lattice Field Theory - LATTICE 2013\\
		July 29 - August 3, 2013\\
		Mainz, Germany}

\begin{document}

\section{Introduction: HAL QCD approach to nuclear force}
To understand hadronic interactions such as nuclear forces from the fundamental theory, Quantum Chromodynamics (QCD), non-perturbative methods such as the lattice QCD combined with numerical simulations 
are required, since the running coupling constant in QCD becomes large at hadronic scale.
Conventionally the finite size method\cite{Luscher:1990ux} has been employed to extract the scattering phase shift
in lattice QCD, but the method is mainly applied to two-particle systems below the inelastic threshold so far.

\begin{figure}[tbh]
\begin{center}
  \includegraphics[width=0.5\textwidth]{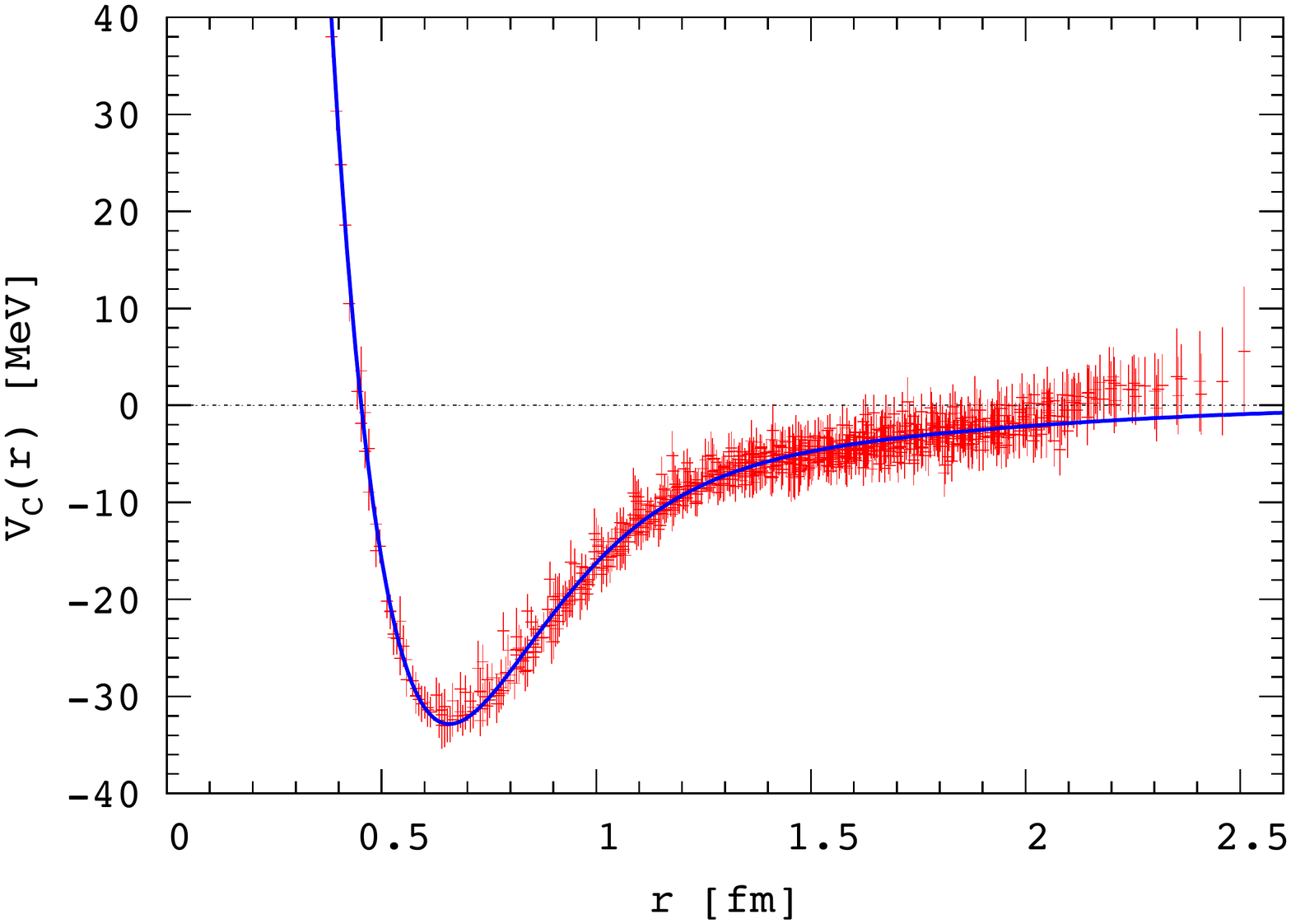}\hfill
  \includegraphics[width=0.5\textwidth]{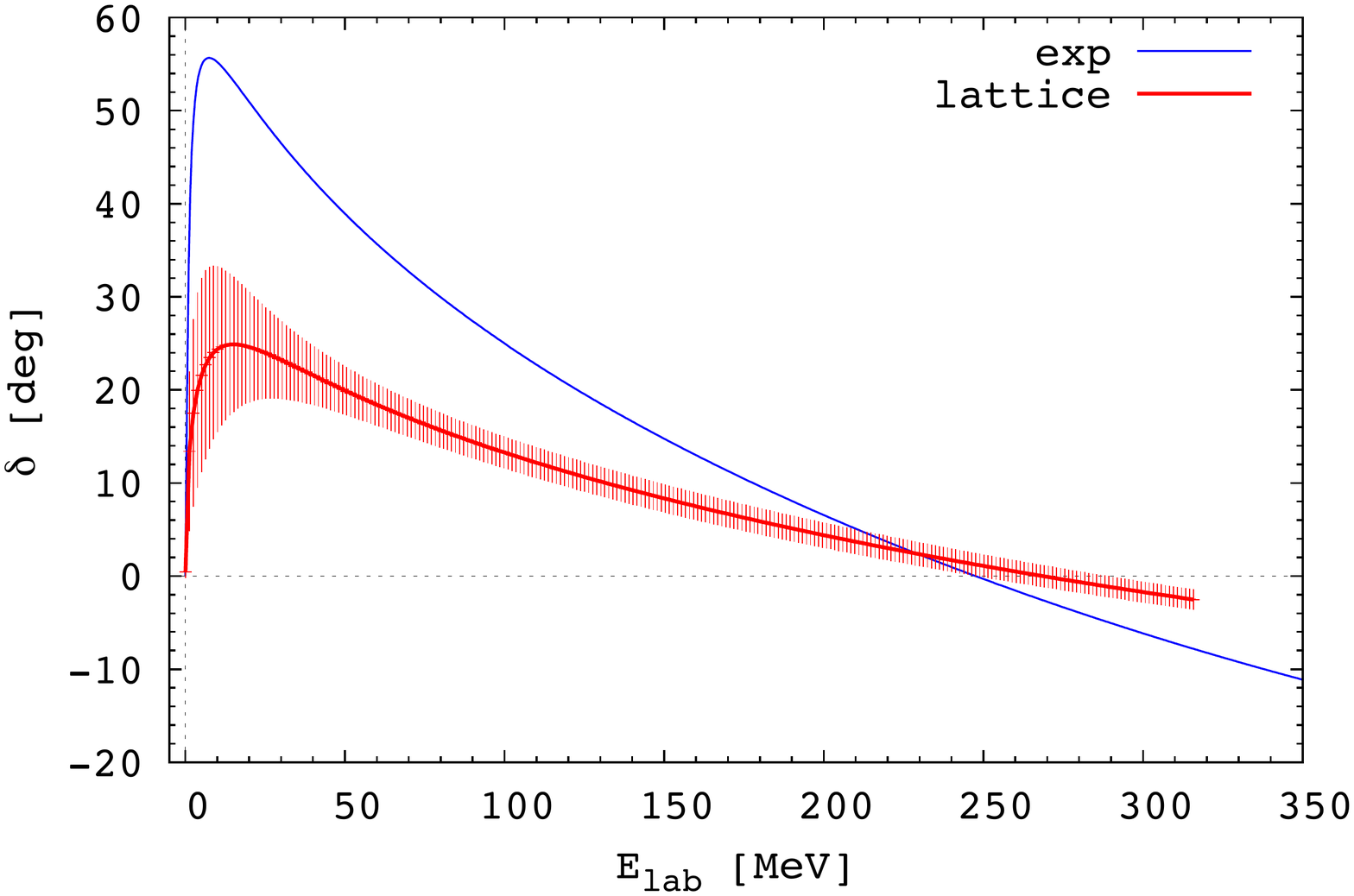}
\caption{(Left) The iso-triplet NN central potential $V_C$ at the leading order of the derivative expansion as a function of $r=\vert\vec{x}\vert$. The solid line is the multi-Gaussian fit of the potential. (Right)  The scattering phase shift in $^1S_0$ channel in the laboratory frame obtained for the potential in the left figure, together with experimental data\cite{nn-online}. Both figures are taken from Ref.~\cite{HALQCD:2012aa}. }
\label{fig:potential}
\end{center}
\end{figure}

Recently an alternative method has been proposed  and  employed to extract the potential between nucleons  below inelastic thresholds\cite{Ishii:2006ec, Aoki:2008hh, Aoki:2009ji,Aoki:2012tk}.
In the method, called the HAL QCD method, a potential between nucleons is defined in QCD, through an equal-time Nambu-Bethe-Salpeter (NBS) wave function\cite{Balog:2001wv} in the center of mass system,  which is defined as
\begin{eqnarray}
\Psi_{\vec{k}}(\vec{x}) &=& \langle 0 \vert T\left\{N(\vec{r}, 0) N(\vec{r} + \vec{x}, 0) \right\}\vert NN,W_k\rangle_{\rm in}
\end{eqnarray}
where $\langle 0 \vert={}_{\rm out}\langle 0 \vert={}_{\rm in}\langle 0 \vert$ is the QCD vacuum (bra-)state,
$ \vert NN, W\rangle_{\rm in} $ is a two-nucleon asymptotic in-state at the total energy $W_k=2\sqrt{\vec{k}^2+m_N^2}$ with the nucleon mass $m_N$ and a relative momentum $\vec{k}$,  $T$ is the time-ordered product, and $N(x)$ with $x=(\vec{x},t)$ is a nucleon operator. 
As the distance between two nucleon operators, $x= \vert \vec{x} \vert$, becomes large, the NBS wave function satisfies the free Schr\"odinger equation, 
\begin{eqnarray}
\left( E_k - H_0 \right) \Psi_{\vec{k}}(\vec{x}) &\simeq& 0, \qquad  E_k=\frac{\vec{k}^2}{2\mu},\quad H_0= \frac{-\nabla^2}{2\mu}, 
\label{eq:asymptotic}
\end{eqnarray}
where $\mu = m_N/2$ is the reduced mass. In addition, an asymptotic behavior of the NBS wave function is described in terms of the phase $\delta$ determined by the unitarity of the $S$-matrix in QCD as\cite{Lin:2001ek, Aoki:2005uf,Ishizuka2009a}
\begin{eqnarray}
\Psi_k^L(x) &\simeq& A_L \frac{\sin ( k x - L\pi/2+\delta_L(W) )}{k x}, \qquad k=\vert\vec{k}\vert, 
\label{eq:phase}
\end{eqnarray}
at $W\le W_{\rm th} = 2m_N+ m_\pi$ (the lowest inelastic threshold),
for the partial wave with the orbital angular momentum $L$. Form this property, it is natural to define a non-local but energy-independent potential $U(\vec{x},\vec{y})$ as
\begin{eqnarray}
\left( E_W - H_0 \right) \Psi_{\vec{k}}(\vec{x}) &=& \int d^3 y\, U(\vec{x},\vec{y}) \Psi_{\vec{k}}(\vec{y}),
\end{eqnarray}
where $U$ does not depend on the energy $W_k$ of a particular NBS wave function.
It is not so difficult to construct such $U$ explicitly as
\begin{eqnarray}
U(\vec{x},\vec{y}) &=& \sum_{\vec{k},\vec{k^\prime}}^{W_k, W_{k^\prime} < W_{\rm th}} (E_{k} - H_0)\Psi_{\vec{k}}(\vec{x}) \eta_{\vec{k},\vec{k^\prime}}^{-1} \Psi^{\dagger}_{\vec{k^\prime}}(\vec{y}),
\end{eqnarray}
where $\eta_{\vec{k},\vec{k^\prime}}^{-1}$ is the inverse of 
\begin{eqnarray}
\eta_{\vec{k},\vec{k^\prime}} = (\Psi_{\vec k}, \Psi_{\vec{k^\prime}}) \equiv
\int d^3 x\, \Psi_{\vec k}^\dagger (\vec{x}) \Psi_{\vec{k^\prime}}(\vec{x})
\end{eqnarray}
in the space spanned by $\{ \Psi_{\vec{k}}\ ,\ W_k < W_{\rm th} \}$. Indeed we see
\begin{eqnarray}
\int d^3y\, U(\vec{x},\vec{y}) \Psi_{\vec{p}}(\vec{y}) &=&\sum_{\vec{k},\vec{k^\prime}}^{W_k, W_{k^\prime} < W_{\rm th}} (E_{k} - H_0)\Psi_{\vec{k}}(\vec{x}) \eta_{\vec{k},\vec{k^\prime}}^{-1} \eta_{\vec{k^\prime},\vec{p}} 
=(E_{p} - H_0)\Psi_{\vec{p}}
\end{eqnarray}
for $^\forall\vec{p}$ with $W_p < W_{\rm th}$. This construction is formal, while in practice we introduce the derivative expansion that
$U(\vec{x},\vec{y}) = V(\vec{x},\vec\nabla)\delta^{(3)}(\vec{x}-\vec{y})$, and the leading order potential can be obtained simply as
\begin{equation}
V_{\rm LO}(\vec{x}) = \frac{(E_k - H_0)\Psi_{\vec{k}}(\vec{x})}{\Psi_{\vec{k}}(\vec{x})} .
\end{equation}
Once the potential is obtained, we can calculate physical observables such as scattering phase shifts and  binding energies, by solving the corresponding Schr\"odinger equation  in the infinite volume.
As such an example, we give the NN potential in the iso-triplet channel obtained in 2+1 flavor QCD at $a\simeq 0.09$ fm, $L a \simeq 2.9$ fm and $m_\pi \simeq 700$ MeV, and the $^1S_0$ scattering phase shift  in Fig.~\ref{fig:potential}\cite{HALQCD:2012aa}, together with experimental data of the phase shift.
The potential as well as the scattering phase shift well reproduce the qualitative feature of the nuclear force, though the attraction at low energy is still weaker than that in nature, probably due to the heaver pion mass. 

In the above proposal so far, considerations are restricted to elastic scatterings of two particles.
In order to extend the HAL QCD method to inelastic and/or multi-particle scatterings, we have to derive two important properties, one is  an asymptotic behavior of NBS wave functions for more than 2 particles\cite{Aoki:2013cra}, and the other is an existence of energy-independent potentials above inelastic thresholds\cite{Aoki:2012bb}.  In this report, the first property is shown in Sec.~\ref{sec:asymptotic},  while the second is discussed in  Sec.~\ref{sec:E-indep}. Our conclusion is given in Sec.~\ref{sec:conclusion}.

\section{NBS wave functions for multi-particles}
\label{sec:asymptotic}
In this section, for simplicity, we consider $n$ scalar particles, all of which have the same mass $m$ but different flavors, denoted by $i$. We also assume that no bound state exists in this system.
\subsection{Unitarity constraint}
In order to derive an asymptotic behavior of NBS wave functions for multi-particles, we first determine structures of $T$ matrix using the unitarity constraint that
\begin{eqnarray}
T^\dagger - T &=& i T^\dagger T , \quad \langle [{\vec p^A}]_n \vert T \vert [{\vec p^B}]_n\rangle =\delta(W^A-W^B)\delta^{(3)}(\vec{P}^A-\vec{P}^B) T(\vec{Q}_A, \vec{Q}_B)
\end{eqnarray}
where $[\vec{p}^{X}]_n=\{\vec{p}^{X}_1, \vec{p}^{X}_2,\cdots, \vec{p}^{X}_n \}$ are momenta of $n$ particles, $W^X =\sum_{i=1}^n W_i^X$ is a total energy with $W_i^X=\sqrt{(\vec{p}_i^X)^2+m^2}$ , and $\vec{P}^X=\sum_{i=1}^n \vec{p}_i^X$ is a total momentum for $X=A,B$, while
we introduce modified Jacobi coordinates and momenta as
\begin{eqnarray}
\vec r_k &=&\sqrt{\frac{k}{k+1}}\left(\frac{1}{k}\sum_{i=1}^k\vec{x}_i - \vec{x}_{k+1}\right),\quad \vec{q}_k =\sqrt{\frac{k}{k+1}} \left(\frac{1}{k}\sum_{i=1}^k\vec{p}_i - \vec{p}_{k+1}\right) ,
\end{eqnarray}
for $k=1,2,\cdots, n-1$, and $\vec{Q}_X=( \vec{q}^{X}_1, \vec{q}^{X}_2,\cdots, \vec{q}^{X}_{n-1} )$.
Regarding $\vec{Q}_X$ as a momentum in $D=3(n-1)$ dimensions, we expand $T$ as
\begin{eqnarray}
T(\vec{Q}_A,\vec{Q}_B) &=& \sum_{[L],[K]} T_{[L][K]}(Q_A,Q_B) Y_{[L]}(\Omega_{\vec{Q}_A}) \overline{ Y_{[K]}(\Omega_{\vec{Q}_B})},\quad Q_A=\vert \vec{Q}_A\vert,  Q_B=\vert \vec{Q}_B\vert,  
\end{eqnarray}
where $ Y_{[L]}$ is the hyper-spherical harmonic function with a set of "quantum" number $[L]=L, M_1, M_2,\cdots$, which satisfies $\hat L^2 Y{[L]}(\Omega_{\vec{s}}) = L(L+D-2) Y_{[L]} (\Omega_{\vec{s}})$ for the  orbital angular momentum $\hat L$ and the solid angle $\Omega_{\vec{s}}$ of the  vector $\vec{s}$ in $D$-dimensions. 

With the non-relativistic approximation that $W^A - W^B \simeq (Q_A^2-Q_B^2)/(2m)$ and the orthogonal property of $Y_{[L]}$ such that
\begin{eqnarray}
\int d\Omega_{\vec{s}}\, \overline{Y_{[L]}(\Omega_{\vec{s}})}\, Y_{[K]}(\Omega_{\vec{s}}) = \delta_{[L][K]}, 
\end{eqnarray}
a solution to the unitarity constraint is given by
\begin{eqnarray}
T_{[L][K]}(Q,Q) &=&\sum_{[N]} U_{[L] [N]}(Q) T_{[N]}(Q) U^\dagger_{[N] [K]}(Q), 
\end{eqnarray}
whose diagonal part  becomes
\begin{eqnarray}
T_{[L]}(Q) &=& - \frac{2n^{3/2}}{m Q^{3n-5}}e^{i\delta_{[L]}(Q)}\sin \delta_{[L]}(Q)
\end{eqnarray}
with a "phase shift" $\delta_{[L]}(Q)$\cite{Aoki:2013cra}.
At $n=2$, the above result reproduces the expression of $T$ for the two particle scattering in the non-relativistic limit that
\begin{equation}
U_{[L]}(Q)=1, \quad T_{[L]}(Q) =T_L(Q) = -\frac{4\sqrt{2}}{mQ}e^{i\delta_L(Q)}\sin \delta_L(Q)
\end{equation}
with $\vec{Q} =(\vec{p}_1-\vec{p}_2)/\sqrt{2}$.

\subsection{Asymptotic behavior of NBS wave functions}
The Lippmann-Schwinger equation in quantum field theories reads
\begin{eqnarray}
\vert \alpha\rangle_{\rm in} &=& \vert  \alpha\rangle_0 +\int d\beta\, \frac{\vert \beta\rangle_0 T_{\beta\alpha}}{E_\alpha-E_\beta + i\varepsilon},
\end{eqnarray}
where the asymptotic state $\vert \alpha\rangle_{\rm in}$ satisfies $ (H_0 + V) \vert \alpha\rangle_{\rm in} = E_\alpha \vert \alpha\rangle_{\rm in}$, while the free state satisfies $ H_0 \vert \alpha\rangle_0 = E_\alpha \vert \alpha\rangle_0$, and off-shell $T$ matrix $T_{\beta\alpha} = {}_0\langle\beta\vert V \vert \alpha\rangle_{\rm in}$  is related to the on-shell $T$-matrix as ${}_0\langle\beta\vert T \vert\alpha\rangle_0 = 2\pi\delta(E_\beta-E_\alpha) T_{\beta\alpha}$\cite{text_weinberg}.

Using the Lippmann-Schwinger equation, 
the NBS wave function, defined by $\Psi_\alpha^n([\vec{x}]) =\langle 0\vert \varphi^n([\vec{x}],0)\vert \alpha \rangle_{\rm in}$ with $\varphi^n([\vec{x}], 0)\equiv T\left\{\prod_{i=1}^n \varphi_i(\vec{x}_i,t)\right\}$,  can be expressed as
\begin{eqnarray}
Z_{\alpha} \Psi^n_\alpha([\vec{x}]) &=&{}_{0}\langle 0\vert \varphi^n([\vec{x}],0)\vert \alpha \rangle_{0}
+\int d\beta\,\frac{Z_\beta}{Z_\alpha}\frac{{}_{0}\langle 0\vert \varphi^n([\vec{x}],0)\vert \beta \rangle_{0} T_{\beta\alpha}}{E_\alpha-E_\beta +i\varepsilon},
\end{eqnarray}
where $Z_\alpha$ is a renormalization factor due to the difference between $\langle 0\vert$ and ${}_{0}\langle 0\vert$, and we can write ${}_{0}\langle 0\vert \varphi^n([\vec{x}],0)\vert \alpha \rangle_{0} = \prod_{i=1}^n e^{i\vec{p}_i\cdot\vec{x}_i}/\sqrt{(2\pi)^3 2E_{p_i}}$.

Expressing $\vec{x}_i$ and $\vec{p}_i$ in terms of modified Jacobi coordinates $\vec{r}_i$ and momenta $\vec{q}_i$, the above equation can be written in the $D$-dimensional hyper-coordinate as
\begin{eqnarray}
Z(\vec{Q}_A) \Psi^n(\vec{R},\vec{Q}_A) &=& \left[e^{i\vec{Q}_A\cdot \vec{R}} +\frac{2m}{2\pi n^{3/2}Z(\vec{Q}_A)}\int d^DQ\,
\frac{Z(\vec{Q}) e^{i\vec{Q}\cdot \vec{R}}}{Q_A^2-Q^2+i\varepsilon}T(\vec{Q},\vec{Q}_A)\right],  
\label{eq:NBS}
\end{eqnarray}
where
$R=(\vec{r}_1,\vec{r}_2,\cdots,\vec{r}_{n-1})$.

As in the previous subsection,  in terms of the hyper-spherical harmonic function, we can write
\begin{eqnarray}
e^{i\vec{Q}\cdot\vec{R}} &=& (D-2)!!\frac{2\pi^{D/2}}{\Gamma(D/2)}\sum_{[L]} i^L j_L^D(QR) Y_{[L]}(\Omega_{\vec{R}}) \overline{ Y_{[L]}(\Omega_{\vec{Q}})},\quad R=\vert\vec{R}\vert, \\
Z(\vec{Q}_A) \Psi^n(\vec{R},\vec{Q}_A) &=& \sum_{[L],[K]} \Psi^n_{[L],[K]}(R,Q_A) Y_{[L]}(\Omega_{\vec{R}})\overline{ Y_{[K]}(\Omega_{\vec{Q}_A})},
\end{eqnarray}
where $j_L^D(x)$ is the hyper-spherical Bessel function.
In the large $R$ limit, since $Q$ integral in eq.~(\ref{eq:NBS}) is dominated at the pole from the denominator, an asymptotic behavior of the NBS wave function now becomes\cite{Aoki:2013cra}
\begin{eqnarray}
\Psi^n_{[L],[K]}(R,Q_A) &\simeq & i^L \frac{(2\pi)^{D/2}}{(Q_A R)^{\frac{D-1}{2}}}\sum_{[N]} U_{[L][N]}(Q_A) e^{i\delta_{[N]}(Q_A)} U^\dagger_{[N][K]}(Q_A)\nonumber \\
&\times& 
\sqrt{\frac{2}{\pi}}\sin\left(Q_A R -\Delta_L + \delta_{[N]}(Q_A) \right), \quad \Delta_L =\frac{2L+D-3}{4}\pi ,
\end{eqnarray}
which demonstrates that we can obtain the $n$ particle scattering "phase shift" $\delta_{[N]}(Q_A)$, determined  by the unitarity constraint of the $T$-matrix,  from the NBS wave function at large $R$.  
Thus the fist property necessary to generalize the HAL QCD method for multi-particle scatterings has been shown.
This property can be shown also for more general inelastic $n_1\rightarrow n_2$ scatterings with $n_1\not= n_2$\cite{Aoki:2013cra}.  
 
\section{Energy-independent potential above inelastic thresholds}
\label{sec:E-indep}
At energy $W > W_{\rm th}$, inelastic $NN\rightarrow NN+ n \pi$ scatterings can occur in addition to the elastic $NN\rightarrow NN$ scattering. To make our discussion simpler, let us consider the $n=1$ case, which corresponds to an energy interval that $m_\pi < W -2m_N <  2 m_\pi $.
In this case, 4 NBS wave functions can be defined as
\begin{eqnarray}
Z_N \varphi^{00}_{W,c_0}(\vec{x}_0) &=& \langle 0 \vert T\{ N(\vec{x},0) N(\vec{x}+\vec{x}_0,0)\} \vert NN,W,c_0\rangle_{\rm in}, \\
Z_N Z_\pi^{1/2} \varphi^{10}_{W,c_0}(\vec{x}_0,\vec{x}_1) &=& \langle 0 \vert T\{ N(\vec{x},0) N(\vec{x}+\vec{x}_0,0)\pi(\vec{x}_1,0)\} \vert NN,W,c_0\rangle_{\rm in},  \\
Z_N \varphi^{01}_{W,c_1}(\vec{x}_0) &=& \langle 0 \vert T\{ N(\vec{x},0) N(\vec{x}+\vec{x}_0,0)\} \vert NN+\pi,W,c_1\rangle_{\rm in}, \\
Z_N Z_\pi^{1/2} \varphi^{11}_{W,c_1}(\vec{x}_0,\vec{x}_1) &=& \langle 0 \vert T\{ N(\vec{x},0) N(\vec{x}+\vec{x}_0,0)\pi(\vec{x}_1,0)\} \vert NN+\pi,W,c_1\rangle_{\rm in},
\end{eqnarray}
where $Z_N$, $Z_\pi$ are renormalization factors, and $c_0,c_1$ denote quantum numbers other than the energy $W$.  For simplicity we express these NBS wave functions as $\varphi^{ij}_{W,c_j}([\vec{x}]_i)$ with $i,j = 0,1$, where $i(j)$ denotes a number of $\pi$'s in the operator(state), and $[\vec{x}]_0=\vec{x}_0$,
$[\vec{x}]_1=\vec{x}_0, \vec{x}_1$

In Ref.~\cite{Aoki:2012bb}, we have shown an existence of non-local but energy-independent potential matrix which satisfies the following coupled channel Schr\"odinger equation: 
\begin{eqnarray}
(E^k_W - H_0^k) \phi^{ki}_{W,c_i} ([\vec{x}]_k)&=& \sum_{l=0,1}\int \prod_{n=0}^ld^3y_n\, U^{kl}([\vec{x}]_k, [\vec{y}]_l)
\varphi^{li}_{W,c_i}([\vec{y}]_l)
\end{eqnarray}
for $k,i = 0,1$, where $E^k_W$ and $H_0^k$ are the kinetic energy and the free Hamiltonian for $NN + k\pi$, respectively. Note that, to show the existence of  $U^{kl}([\vec{x}]_k, [\vec{y}]_l)$, we have used a non-relativistic approximation that
\begin{equation}
W - 2m_N - k m_\pi \simeq E_W^k = \frac{\vec{p}_1^2}{2m_N} + \frac{\vec{p}_2^2}{2m_N} +\sum_{i=1}^k \frac{\vec{k}_i^2}{2m_\pi}, \quad \vec{p_1}+\vec{p}_2+\sum_{i=1}^k \vec{k}_i = 0,
\end{equation}
in the center of mass frame. See Ref.~\cite{Aoki:2012bb} for an explicit construction of $U^{kl}([\vec{x}]_k, [\vec{y}]_l)$.

The above construction of non-local but energy-independent potential matrix $U$ can be easily generalized to $NN+n_1\pi \rightarrow NN+n_2\pi$\cite{Aoki:2012bb}  or to $ \Lambda\Lambda \rightarrow \Lambda\Lambda, N\Xi, \Sigma\Sigma$\cite{Aoki:2011gt,
Aoki:2012bb} 

\section{Conclusion}
\label{sec:conclusion}
The HAL QCD approach is a promising method to extract hadronic interaction in lattice QCD. 
In this conference, there appear several new results in this approach, which include a LS nuclear force\cite{Murano:2013xxa}, an anti-symmetric LS force between octet baryons in the flavor SU(3) limit\cite{ishii}, interactions of DD$^{*}$, $\overline{\rm K}$D and $\overline{\rm K}$D$^{*}$\cite{ikeda}, the Omega-Omega interactions\cite{yamada}.
Comparisons between the HAL QCD method and the L\"uscher's finite volume method have been also reported for the NN case\cite{charon} and the $\pi\pi$ case\cite{Kurth:2013tua}.

In this report, we have attempted an extension of the HAL QCD method to inelastics and/or multi-particle scatterings, by determining asymptotic behaviors of NBS wave functions for $n$ scalar particles in terms of the generalized scattering phase of the $T$-matrix, and by constructing non-local but energy-independent coupled channel potentials for inelastic scatterings.
Our results are relevant for the 3 nuclear force\cite{Doi:2011gq} and the coupled channel potentials between hyperons\cite{sasaki}.   

One of remaining issues in this study is a treatment of bound-states. For example, in the three nucleon scattering, two of them could become a bound deuteron.  Whether a bound state is treated as an independent state or not is an open question in the extension of the HAL QCD approach.
We will address this issue in our future publications.\\

This work is supported in part by the
Grant-in-Aid for Scientific Research (25287046), the Grant-in-Aid for Scientific Research on Innovative Areas(No.2004: 20105001, 20105003) and SPIRE (Strategic Program for Innovative Research).

\end{document}